# Correlation Between Phase Competition and the Nucleation of a Griffiths Phase in $(La_{1-y}Pr_y)_{0.7}Ca_{0.3}Mn^{16/18}O_3$

Wanjun Jiang, XueZhi Zhou, and Gwyn Williams
Department of Physics and Astronomy, University of Manitoba, Winnipeg MB R3T 2N2 CANADA

Detailed analyses of the temperature-dependent zero field ac susceptibility of prototypical phase-separated $(La_{1-y}Pr_y)_{0.7}Ca_{0.3}Mn^{16/18}O_3$, $0 \leq y \leq 1$, reveal features consistent with the presence of a Griffiths phase (GP), viz., an inverse susceptibility ($\chi^{-1}$) characterized by $\chi^{-1} \propto (T - T_C^{Rand})^{1-\lambda}$ with $0.05 \leq \lambda \leq 0.33$ as y decreases towards $y_c \leq 0.85$. Beyond $y_c = 0.85$, the GP is suppressed. These data, combined with previous neutron diffraction measurements, enable a phase diagram summarizing the evolution of the GP with composition to be constructed for this system; in particular, it shows that the disorder relevant for the establishment of such a phase is linked closely to the relative volume fractions of the phase separated antiferromagnetic and ferromagnetic components, even when the recently estimated double exchange (DE) linked percolation threshold is exceeded. The influence of electron-phonon coupling can also be seen through oxygen isotope effects.

Colossal magnetoresistive (CMR) materials constitute an important class of strongly correlated electronic systems displaying an unusual combination of coupling between charge, spin, orbital and vibrational degrees of freedom. One, possibly the, important characteristic of these materials is that the ordering tendencies associated with these degrees of freedom are frequently in competition, producing nearly degenerate ground states with markedly different characteristics [1-5]. A corollary to such complexity is that the role played by quenched disorder has emerged as pivotal in elucidating the behaviour of these materials [1-6].

Perovskite manganite materials exhibiting CMR are often characterized by the general formula $A_{1-x}B_xMnO_3$ (A = rare-earth while B = a divalent alkaline earth ion), x being the doping level which changes the valence state of the Mn ions to maintain charge neutrality (hence modulating the extent of "double-exchange" (DE) in models of CMR). The consequent mismatch in the ionic size of ions occupying the "A" (La) site leads to a distortion in the perovskite structure, variously characterised by the Goldschmidt tolerance factor, $t$, the average A-site radius, $<r_A>$, and the variance, $\sigma$, in the A-site radius. While these latter parameters provide quantitative measures of the static structural disorder in these systems [6], ionic size mismatch also modifies the Mn-O-Mn bond lengths and angles, and consequently both DE and super-exchange coupling [6]. These latter examples of disorder are unavoidable consequences of doping in manganites, in addition to the obvious site disorder.

Recently, the observation of a Griffiths phase (GP) [7] in number of doped perovskite manganites has been reported [8-13]. While the relationship between the presence of such a phase and the onset of CMR is the subject of ongoing discussion, the nucleation of this phase is linked unequivocally to the presence of quenched disorder. In the context of the manganites, one pivotal question outstanding is whether there is one particular "measure" of disorder which characterises the appearance of a GP [12]. In an attempt to address this question, we present an analysis of systematic measurements of ac susceptibility on a series of fixed, "optimally" hole doped $(La_{1-y}Pr_y)_{0.7}Ca_{0.3}Mn^{16/18}O_3$ ($0 \leq y \leq 1$) polycrystalline compounds [14]. The use of such samples importantly maintains not only a fixed number of DE linked $Mn^{3+}$-$Mn^{4+}$ sites, but also a fixed number of $Mn^{3+}$-$Mn^{3+}$ and $Mn^{4+}$-$Mn^{4+}$ super-exchange interactions (the possible influence of substitution on the magnitude and sign of these interactions notwithstanding). Such an analysis, used in conjunction with previous neutron powder diffraction data, indicates first that there is a close correlation between phase competition as measured by the relative volume fractions of the phase separated antiferromagnetic (AFM) and ferromagnetic (FM) components, and the nucleation of a GP, and second, the GP in this system resembles that conjectured for the ±J random bond Ising model (RBIM) [7, 8], in that it is confined to a restricted regime of the temperature (T)-probability (p) plane, p here being related to the relative FM/AFM volume ratio.

In Griffiths' original treatment of a diluted FM Ising ferromagnets [7], nearest neighbor exchange bonds of strength J occurred with probability p, while bonds of zero strength and probability (1-p) represented disorder. Above the percolation threshold, $p_c$, of the relevant lattice FM order is established, but, as expected, at a temperature $T_C(p)$ below that of the undiluted system [$T_C(p=1) = T_G$]. By contrast, for $p < p_c$, there is zero probability of establishing an infinite percolating "backbone" (so the correlation length does not diverge as in a continuous transition), and thus no cooperative ferromagnetism is established. However in the Griffiths temperature interval $T_C(p) < T < T_G$ the system response is neither simply paramagnetic (PM)/Curie-Weiss (CW)-like nor is the correlation length divergent; in this regime the response is dominated by the largest magnetic cluster/correlated volume, leading to a characteristic temperature dependence for the susceptibility ($\chi$), viz. [7]:

$$\chi^{-1} \propto (T - T_C^{Rand})^{1-\lambda}, 0 \leq \lambda < 0 \qquad (1)$$

This power law has been shown to reproduce the low field ac and dc susceptibilities immediately above $T_C$ in a variety of candidate GP systems [7-13, 15], its form demonstrating clearly the depression of $\chi^{-1}$ in this temperature regime below its CW value, as a result of the formation of large correlated regions/clusters[16]. Here $T_C^{Rand}$, and the transition/Griffiths temperature, $T_G$, of the undiluted system are determined from such data, as outlined below.

The main body of Figs. 1(a-e) present a selection of inverse zero field ac susceptibility data, which clearly monitors the nucleation and evolution of a GP with composition in this optimally doped system. Data from $^{16}$O doped samples are presented on the left and for $^{18}$O doping on the right. Double-logarithmic plots of $\chi^{-1}$ against reduced temperature, $t_m = (T - T_C^{Rand})/T_C^{Rand}$, reproduced in the inserts in these figures, confirm the power-law prediction-Eq. (1)- and yield estimates for the exponent $\lambda$ (notwithstanding the difficulty surrounding such estimates at small $\lambda$ values).

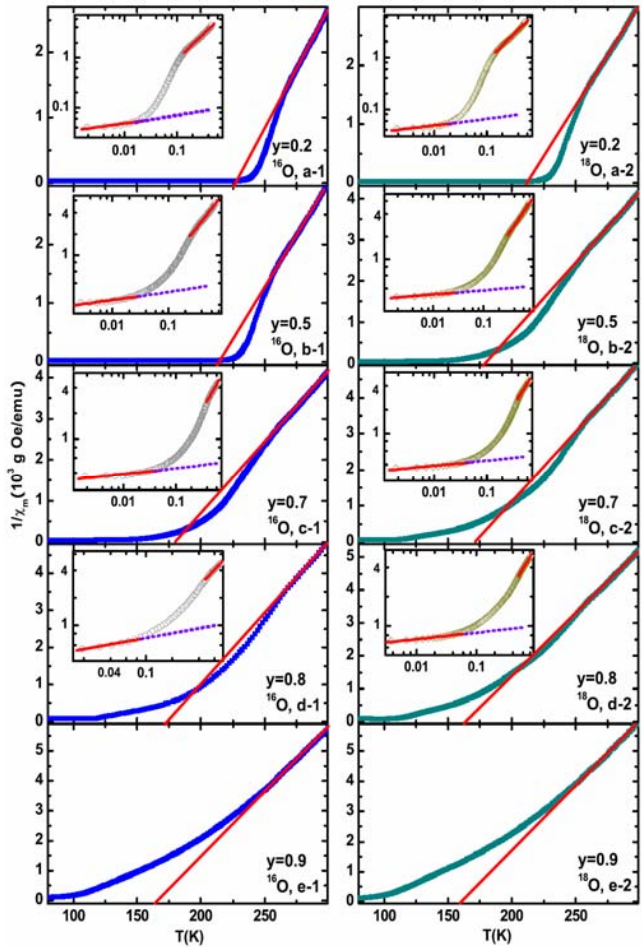

Figure 1: (Color online) Main body: the inverse ac susceptibility against temperature, measured on warming; $^{16}$O and $^{18}$O data are shown on the left and right, respectively. The Pr substitution levels, y, are marked. The inserts replot these data on a double-logarithmic scale, testing the power-law Eq. (1) with reduced temperature $t_m = (T - T_C^{Rand})/T_C^{Rand}$ and yielding estimates for exponent $\lambda$.

While specific criteria exist for determining the PM to FM ordering temperature, $T_C$, and the Griffiths temperature, $T_G$, viz., the inflection point in the zero-field ac susceptibility [14] and the onset of marked departures from CW behaviour [7, 11-13, 15], respectively, the choice of $T_C^{Rand}$ has been less precise. Here we propose a definite criterion, based on several previously reports [8-13, 15], that fitting data in the PM regime above $T_G$ to Eq. (1) yields a value for $\lambda$ close to zero when $T_C^{Rand}$ is correctly identified. Such a result is expected as the GP evolves into a conventionally disordered PM phase. The use of this criterion not only enables $T_C^{Rand}$ to be acquired consistently, but it also avoids the unacceptable occurrence of both negative values for $\lambda$ when choices of $T_C^{Rand}$ below its correct value are adopted, and increasingly positive estimates for $\lambda$ which emerge when too high a $T_C^{Rand}$ is used in Eq. (1) (trends that are exacerbated in self-consistent approaches). The associated accuracy in the estimates for $T_C^{Rand}$ using this criterion is ±1 K, as shown in Table 1 below. Table 1 also lists the corresponding values for $T_G$ ($T_G \approx 265$ K for $^{16}$O, and $\approx 260$ K for $^{18}$O, essentially constant for all samples exhibiting GP except close to the critical composition, $y_c = 0.85$ in $^{16}$O), $\lambda$ and $T_C$, parameters used subsequently to construct Fig. 2a, and 2b.

Irrespective of any detailed analysis based on Eq. (1), the clear observation of a depression of $1/\chi$ below its projected high-temperature CW form in Figs. 1 a1 - d1 provides unequivocal evidence for the formation of large clusters/correlated regions [16], consistent with the formation of a GP. Also clearly evident from Figs. 1e is that higher levels of Pr substitution (y = 0.9) suppresses the nucleation of such clusters, and correspondingly, the GP. The immediate implication is that $y_c = 0.85$ represents the compositional critical point below which a GP is nucleated in this system (comparable results occur with $^{18}$O substitution, Figs 1 a2-e2, a point returned to below). The idea of a critical ratio in a phase-separated view of the manganites has been discussed recently [17], focusing principally on the thermal, not compositional, variation of this ratio. By contrast, and of immediate relevance to the present conclusion, is that analysis of previous neutron diffraction experiments [14] concluded that at this same composition the role of quenched disorder was pivotal in establishing a long-ranged phase separated state. The significance of this correlation is emphasized by an inspection of Figs. 2a-2d. The first of these reproduce the three characteristic temperatures $T_C$, $T_C^{Rand}$ and $T_G$ as a function of La replacement by Pr (y) (while maintaining *optimal* doping) in both $^{16}$O (Fig. 2a) and $^{18}$O systems (Fig. 2b); these phase diagrams exhibit a trapezoidal shape, similar to that first reported in La$_{1-x}$Sr$_x$MnO$_3$ ([8], and also La$_{1-x}$Ba$_x$MnO$_3$ [12]), which replicates aspects of the conjectured phase diagram for the FM regime of the RBIM. These figures also emphasize the precipitous drop in $T_G$ beyond $y_c = 0.85$ where a GP fails to nucleate, again inviting comparison with the AFM region of the RBIM.

Table 1: Static structural parameters characterizing $(La_{1-y}Pr_y)_{0.7}Ca_{0.3}Mn^{16/18}O_3$ ($0 \leq y \leq 1$). The ionic radii are from [20], the y=0 data cited from [10]. "--" indicates no data available at that composition.

| y | $\langle r_A \rangle$ (Å) | $\sigma$ (Å) | t tolerance factor | $T_C$ (K) -$^{16}$O | $T_C$ (K) -$^{18}$O | $T_C^{Rand}$ (K) -$^{16}$O | $T_C^{Rand}$ (K) -$^{18}$O | $T_C(\langle r_A \rangle,0)$ (K) | $T_G$ (K) $^{16}$O | $T_G$ (K) $^{18}$O | $\lambda$-$^{16}$O | $\lambda$-$^{18}$O |
|---|---|---|---|---|---|---|---|---|---|---|---|---|
| 0 | 1.2052 | 0.01803 | 0.9163 | 212 | -- | 216 | -- | 270 | 245 | -- | 0.33 | -- |
| 0.2 | 1.2000 | 0.01803 | 0.9145 | 232 | 219 | 232 | 219 | 240 | 265 | 260 | 0.15 | 0.12 |
| 0.5 | 1.1923 | 0.01743 | 0.9117 | 180 | 155 | 200 | 188 | 193 | 264 | 258 | 0.13 | 0.09 |
| 0.7 | 1.1871 | 0.01492 | 0.9099 | 138 | 122 | 182 | 171 | 160 | 265 | 259 | 0.09 | 0.08 |
| 0.75 | 1.1858 | 0.0135 | 0.9094 | 122 | 116 | 178 | 167 | 152 | 265 | 256 | 0.08 | 0.07 |
| 0.8 | 1.1845 | 0.01273 | 0.9010 | 120 | 107 | 175 | 166 | 143 | 267 | 255 | 0.07 | 0.05 |
| 0.85 | 1.1832 | 0.01176 | 0.9102 | 116 | -- | 178 | -- | 134 | 228 | -- | 0.05 | -- |
| 0.9 | 1.1819 | 0.00937 | 0.9081 | 99 | 94 | 0 | 0 | 126 | 0 | 0 | 0 | 0 |
| 1 | 1.1793 | 4.58E-4 | 0.9072 | 120 | 115 | 0 | 0 | 108 | 0 | 0 | 0 | 0 |

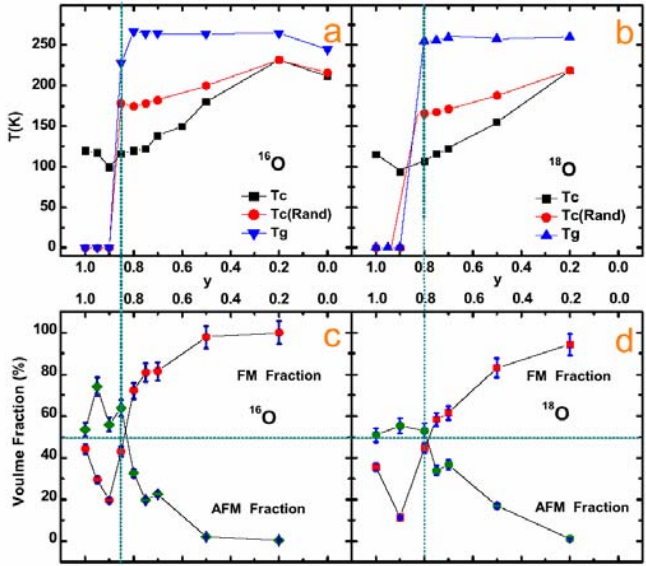

Figure 2: (Color online) Plots of $T_G$, $T_C$, and $T_C^{Rand}$ vs. the Pr doping level, y, for the $^{16}$O (a) and the $^{18}$O substituted series (b). The associated AFM and FM volume fractions calculated as described in the text, in the $^{16}$O (c) and the $^{18}$O series (d).

The consensus of current work is that a GP in the manganites originates from quenched disorder; here we attempt to isolate one principal characteristic of this disorder which can be linked to the nucleation of a GP. The advantages accrued by using a system maintained at optimal doping have been discussed earlier; nevertheless, it should be reiterated that a consequence of changing the La-Pr ratio even at fixed (total) doping is that the Mn-O-Mn bond lengths and angles are inevitably modified. Several previous attempts to provide a unified description of the properties of doped-manganites (and cuprates) [6, 18] have been based on "static" structural factors, particularly $\langle r_A \rangle$, and $\sigma$ mentioned earlier (although changes in both the Mn-O and O-O bonds occur on passing through $T_C$ [19]). Such discussions, however, made no quantitative predictions relevant to the present topic, viz. a comparison between the extent of structural disorder and the magnitude of $T_G$ and/or $T_C^{Rand}$ or the existence of a critical composition in this context. The compendium of structural parameters collected in Table 1 facilitates the reiteration of a previous suggestion [12] that no such parameter-or indeed a combination of them-maps onto the behaviour summarized in Figs 2a-b. In contrast, the present analysis offers a quantitative measure of phase competition and its correlation with the nucleation of a GP. Specifically it demonstrates that the replacement of La by Pr (y) results in progressively decreasing values for the susceptibility "exponent" λ in the GP regime, from a maximum value of λ = 0.33 at y = 0 to λ = 0 above the critical composition, $y_c > 0.85$.

The current analysis utilises previous neutron diffraction data [14]. These data not only demonstrated the presence of competing phases, but also confirmed that an AFM component can be detected at all compositions down to y = 0.2 (at this latter composition the AFM volume fraction is not unexpectedly very small, viz., the effective AFM moment $m_{AFM}$ = 0.24 $\mu_B$ [14]). Here we demonstrate the important result that as the AFM phase fraction increases with the Pr substitution level, y, so the nucleation of a GP is suppressed. The FM/AFM volume fractions were estimated using effective FM ($m_{FM}$) and AFM ($m_{AFM}$) moments measured at low temperature (10 K – 15 K) (briefly, using a pure AFM state moment of $M_{AFM}$ =2.26 $\mu_B$ and its FM counterpart $M_{FM}$ =3.57 $\mu_B$, taken to be constant over the whole Pr (y) doping range, used in conjunction with the relationship: $(m_{AFM}/M_{AFM})^2 = 1-(m_{FM}/M_{FM})^2$, yields the relevant FM volume fraction, $(m_{FM}/M_{FM})^2$, and vice-versa). The resulting FM and AFM volume fractions are reproduced in Figs. 2c ($^{16}$O) and 2d ($^{18}$O) as a function of y, enabling direct comparisons with Figs. 2a and 2b to be made.

This comparison demonstrates a marked correlation linking directly the nucleation of a GP in this system – undoubtedly reflecting the presence of quenched disorder – but disorder quantified by the relative volume fractions of the phase separated FM and AFM components. In particular, the critical Pr composition, $y_c = 0.85$, above which λ vanishes and below which a GP first forms, coincides with the emerging dominance of the FM phase fraction. These figures also demonstrate that in the presence of competition between phase separated AFM and FM regions/clusters, the occurrence of GP is indeed confined to a restricted region of the temperature-probability (T-p) phase diagram, again showing similarities with predictions for the RBIM [8]. Here however, the relevant probability (p) parameter is not simply the A-site occupation, *per se* [8], but the relative AFM/FM volume fraction modulated by Pr doping, with the experimentally determined compositional threshold at $y_c = 0.85$ (equivalent to $p_c = 0.5$ in the RBIM). It is particularly important to note that by maintaining optimal doping with $x = 0.3$, the lower bound of $0.04 < x_c < 0.1$ for the appearance of a GP, estimated recently by Deisenhofer et al., [8] based on the *number* of DE linked sites, is deliberately exceeded throughout the present system; nevertheless, a GP is not ubiquitous, it nucleates only below a Pr probability compositional threshold ($y_c$) where the FM volume fraction dominates. The additional correlation between λ estimates in the GP (once established) and the FM/AMF volume ratio adds support to this assertion, viz. these λ estimates fall monotonically as the Pr concentration approaches $y_c$ and the dominance of the FM component diminishes. Of course, there is a compositional upper bound beyond which a GP fails to form, around x=0.3 in the $La_{1-x}Ca_xMnO_3$ series, as the present data confirm, but for which numerical estimates do not exist at present. The current analysis similarly shows that beyond this supposed upper bound a GP can still be nucleated, with a constant number of DE linked sites, by manipulating the relative FM/AFM volume fraction.

That a Jahn-Teller mediated electron-phonon coupling plays an important role in manganites is confirmed by isotope effects accompanying $^{18}O$ replacement of $^{16}O$ (Figs 1-e1 and e2 and Table 1). Such effects are, however, not dominant here; there is a small depression of some 5 K in $T_G$ (likely due to a decrease in the Mn-O-Mn mode frequencies accompanying the oxygen mass increase [21]), and both $T_C^{Rand}$ and λ are both slightly lower in $^{18}O$ doped specimens. This appears to reflect a similar reduction in the critical composition below $y_c = 0.85$, where, as with its $^{16}O$ counterpart, it again correlates closely with the emerging dominance of the FM phase fraction; the present data indicate that this phase competition is modulated, but certainly not suppressed, by the electron-phonon coupling, as is the accompanying establishment of a GP.

In summary, systematic analysis of the ac susceptibility of a series of $(La_{1-y}Pr_y)_{0.7}Ca_{0.3}Mn^{16/18}O_3$ ($0 \leq y \leq 1$) indicates a compositional critical point $y_c = 0.85$ ($^{16}O$) above which a GP does not nucleate and λ vanishes. Below this critical composition, the characteristic depression of the $\chi^{-1}$ below its high-temperature CW behaviour, summarized by the power law $\chi^{-1} \propto (T - T_C^{Rand})^{1-\lambda}$, $0.05 \leq \lambda \leq 0.33$, is clearly evident. The reanalysis of previous neutron diffraction data [14] demonstrates that for Pr concentrations $y_c > 0.85$, the AFM component in this phase separated manganite dominates; in contrast, the FM component is prevalent below this critical composition, and it is in this latter regime that a GP first appears, despite the fact that all these *optimally* doped samples exceed the lower bound established for the emergence of a GP in DE linked systems. The result that the trend in the "exponent" λ also tracks this FM component suggests that the relative FM-AFM phase fraction in this phase separated system provides the appropriate measure of the disorder from which a GP originates. That there is significant competition between states with differing *magnetic* characteristics (as opposed to disorder measured solely by structurally based parameters) is confirmed, for example, by the observation of a $T_C \geq T_N$ near $y = 0.5$; in particular, the latter demonstrate the presence of two energy scales, the existence of which underlies the appearance of CMR [22] in several computational approaches [23] which rely on such competition to establish intrinsically inhomogeneous ground states. In these strongly correlated systems in which charge, spin, orbital and phononic degrees of freedom are believed to play competing roles, phononic effects, while not dominant, can be seen through oxygen isotope effects.

Support for this work by the Natural Sciences and Engineering Research Council (NSERC) of Canada is gratefully acknowledged. The authors also offer their sincere thanks to Dr. V. Yu. Pomjakushin and colleagues for sending the the raw data on which this reanalysis is based, and their comments on the manuscript.

Electronic address: jiang@physics.umanitoba.ca

Reference:
[1] *Colossal Magneto-resistive Oxides*, edited by Y. Tokura (ISBN 90-5699-231-7); E. Dagotto, *Phase Separation and Colossal Magnetoresistance* (Springer, Berlin, 2002); M. B. Salamon, et al., Rev. Mod. Phys. 73, 583 - 628 (2001).
[2] C. Zener, Phys. Rev. **81**, 440 (1950).
[3] A. J. Millis, et al., Phys. Rev. Lett **74**, 5144 (1995).
[4] A. Moreo, et al., *Science*, **283**, 2034 (1999).
[5] Zhao, Guo-Meng, et al., *Nature*, **381**, 676 (1996).
[6] L. M. Rodriguez-Martinez, et al., Phys. Rev. B **54**, R15622 (1996), Jonker G. H., et al., *Physica* **16** 337 (1950).
[7] R. B. Griffiths, Phys. Rev. Lett **23**, 17 (1969); A. J. Bray, ibid., **59**, 586 (1987); A. H. Castro-Neto, et al., ibid., **81**, 3531 (1998).
[8] J. Deisenhofer, et al., Phys. Rev. Lett **95**, 257202 (2005).
[9] M. B. Salamon, et al., Phys. Rev. Lett **88**, 197203 (2002).
[10] M. B. Salamon, et al., Phys. Rev. B **68**, 014411 (2003).
[11] Wanjun Jiang, et al., Phys. Rev. Lett **99**, 177203 (2007).
[12] Wanjun Jiang, et al., Phys. Rev. B **77**, 064424 (2008).
[13] Wanjun Jiang, et al., Phys. Rev. B **76**, 092404 (2007).
[14] V. Pomjakushin, et al., Phys. Rev. B **75**, 054410 (2007).
[15] C. Magen, et al., Phys. Rev. Lett **96**, 167201 (2006).
[16] J. M. De Teresa, et. al., *Nature* **386**, 256 - 259 (1997).
[17] Z. Sun, et al., Nature Physics, Vol **3**, 248 (2007).
[18] J. P. Attfield, et al., *Nature* **394**, 157 (1998).
[19] S. J. L. Billinge, et al., Phys. Rev. Lett **77**, 715 (1996).
[20] R. D. Shannon, *Acta Cryst,* A**32**, 751 (1976).
[21] F. Rivadulla, et al., Phys. Rev. Lett **96**, 016402 (2006).
[22] M. Uehara, et al., *Nature* (London) **399**, 560 (1999).
[23] A. Moreo, et al., Phys. Rev. Lett **84**, 5568 (2000); J. Burgy, et al., ibid., **87**, 277202 (2001); C. Sen, et al., ibid., **84**, 127202 (2007).